\documentclass[%
 reprint,
 twocolumn,
superscriptaddress,
showpacs,
 amsmath,amssymb,
 aps,
 prc,
]{revtex4}

\usepackage{graphicx}
\usepackage{dcolumn}
\usepackage{bm}
\usepackage[T1]{fontenc}

\begin{document}

\preprint{APS/123-QED}

\title{Pseudo spin doublet bands and Gallagher Moszkowski
doublet bands in $^{100}$Y}

\author{E. H. Wang}
\author{J. H. Hamilton}
\author{A. V. Ramayya}
\author{C. J. Zachary}
 \affiliation{Department of Physics and Astronomy, Vanderbilt University, Nashville, TN 37235, USA}

\author{A. Lemasson}
\author{A. Navin}
\author{M. Rejmund}
 \affiliation{GANIL, CEA/DRF - CNRS/IN2P3, Bd Henri Becquerel, BP 55027, F-14076 Caen Cedex 5, France}
\author{S. Bhattacharyya}
 \affiliation{Variable Energy Cyclotron Centre, 1/AF Bidhan Nagar,
  Kolkata 700064, India}

\author{Q. B. Chen}
\author{S. Q. Zhang}
 \affiliation{State Key Laboratory of Nuclear Physics and Technology, School of Physics,
 Peking University, Beijing 100871, People's Republic of China}

\author{J. M. Eldridge}
\author{J. K. Hwang}
 \affiliation{Department of Physics and Astronomy, Vanderbilt University, Nashville, TN 37235, USA}
\author{N. T. Brewer}
 \altaffiliation{Present address: Physics Division, Oak Ridge National Laboratory, Oak Ridge, TN 37831, USA}
 \affiliation{Department of Physics and Astronomy, Vanderbilt University, Nashville, TN 37235, USA}
\author{Y. X. Luo}
 \affiliation{Department of Physics and Astronomy, Vanderbilt University, Nashville, TN 37235, USA}
 \affiliation{Lawrence Berkeley National Laboratory, Berkeley, CA 94720, USA}
\author{J. O. Rasmussen}
 \affiliation{Lawrence Berkeley National Laboratory, Berkeley, CA 94720, USA}

\author{S. J. Zhu}
 \affiliation{Department of Physics, Tsinghua University, Beijing 100084, People's Republic of China}
\author{G. M. Ter-Akopian}
\author{Yu. Ts. Oganessian}
 \affiliation{Joint Institute for Nuclear Research, RU-141980 Dubna, Russian Federation}

\author{M. Caama\~{n}o}
\affiliation{USC, Universidad de Santiago de Compostela, E-15706
Santiago de Compostela, Spain}

\author{E. Cl\'ement}
\author{O. Delaune}
\author{F. Farget}
\author{G. de France}
\author{B. Jacquot}

\affiliation{GANIL, CEA/DRF - CNRS/IN2P3, Bd Henri Becquerel, BP 55027, F-14076 Caen Cedex 5, France}

\date{\today}

\begin{abstract}
New transitions in neutron rich $^{100}$Y have been identified
in a $^9$Be+$^{238}$U experiment with mass- and Z- gates to provide
full fragment identification. These transitions and high spin levels
of $^{100}$Y have been investigated by analyzing the high statistics
$\gamma$-$\gamma$-$\gamma$ and $\gamma$-$\gamma$-$\gamma$-$\gamma$
coincidence data from the spontaneous fission of $^{252}$Cf at the
Gammasphere detector array. Two new bands, 14 new levels and 23
new transitions have been identified. The $K^{\pi}=4^+$ new band decaying
to an 1s isomeric state is assigned to be the high-$K$
Gallagher-Moszkowski (GM) partner of the known $K^{\pi}=1^+$ band, with
the $\pi 5/2[522] \otimes \nu 3/2[411]$ configuration. This 4$^+$ band is also
proposed to be the pseudo spin partner of the new $K^{\pi}=5^+$ band with
a 5$^{+}$ $\pi 5/2[422] \otimes \nu 5/2[413]$ configuration, to form
a $\pi 5/2[422] \otimes \nu [312$ $5/2,3/2]$ neutron pseudospin
doublet. Constrained triaxial covariant density functional theory
and quantal particle rotor model calculations have been applied to interpret the
band structure and available electromagnetic transition
probabilities and are found in good agreement with experimental
values.
\end{abstract}

\maketitle


\section{\label{sec:level1}Introduction}

In the mass region $A\sim 100$, the shape transition and shape
coexistence of the neutron-rich nuclei have long been of
interest~\cite{Sha,Ham95}. In this region, there exists a shell
closure effect with $Z=40$ and $N=56$ spherical subshells and a
sudden onset of large ground state axially symmetric deformation for
$ N\geq 60$ occurs in the Sr ($Z=38$), Y ($Z=39$) and Zr ($Z=40$)
isotopes \cite{Ham95}. Thus, it is reasonable to consider the
odd-odd $^{100}$Y with $Z=39$ and $N=61$ to have a large quadrupole
deformation.

The pseudo spin is one of the current interests in nuclear study. In
this concept, a pair of single particle orbitals with quantum
numbers ($n-1, l+2, j=l+3/2$) and ($n, l, j=l+1/2$) lie very close
in energy. They can be labeled as a pseudo spin doublet with quantum
numbers ($\tilde{n}=n-1, \tilde{l}=l+1, \tilde{j}=l\pm1/2$). In the
deformed case, the Nilsson states labeled by the asymptotic quantum
numbers $[Nn_z\Lambda]\Omega$ have pseudo-Nilsson equivalents
$[\tilde{N}\tilde{n}_z\tilde{\Lambda}]\tilde{\Omega}$, where
$\tilde{N}=N-1$, $\tilde{n}_z=n_z$, $\tilde{\Lambda}=\Lambda\pm1$,
and $\tilde{\Omega}=\Omega=\tilde{\Lambda}+\tilde{\Sigma}$, where
$\tilde{\Sigma}=\pm1/2$. In the axially symmetric pseudo-Nilsson
scheme, $\tilde{\Lambda}$, $\tilde{\Sigma}$ and $\tilde{\Omega}$ are
good quantum numbers at all deformations~\cite{Stu}. In the A$\sim$
100 region, pseudo spin bands were reported in $^{108}$Tc~\cite{Xu}.

In this paper, we report the new high spin structure of $^{100}$Y
with identification of three new bands and the extension of a
previously reported band~\cite{Woh,Hwa98}. Pseudo spin doublet and
Gallagher-Moszkowski (GM) doublet bands \cite{Gal} are proposed in this
nucleus. The constrained triaxial covariant density functional
theory and quantal particle rotor model calculations were performed
and compared favorably with such assignments.

\section{Experimental method}

\begin{figure*}
 \includegraphics[width=0.94\textwidth]{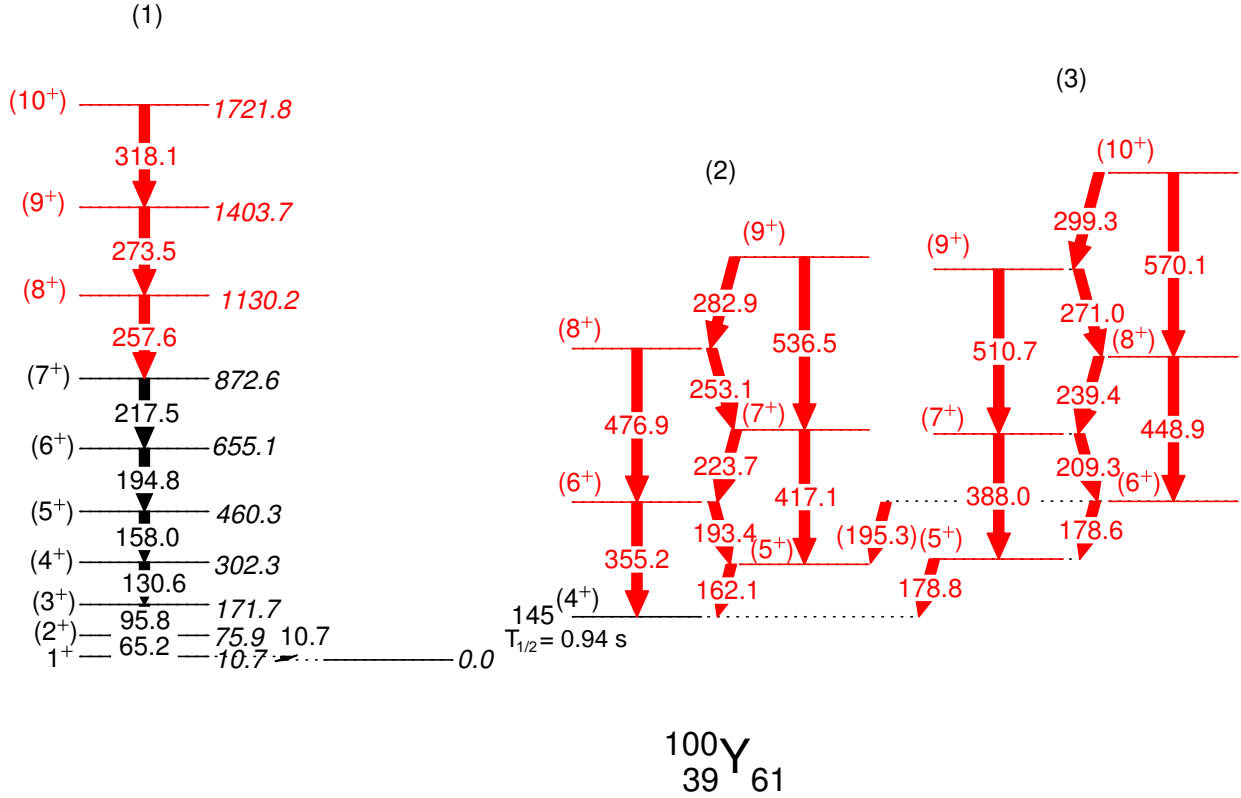}
 \caption{\label{fig1}The new level scheme of $^{100}$Y obtained
in the present work. New transitions and levels are labeled in red. The 10.7 keV transition is not observed in the current work but reported in Ref.~\cite{Woh}. Previous rotational band (1) was reported in Ref.~\cite{Hwa98}. The 145 keV isomer was reported in Ref.~\cite{Hag,Bac}.}
\end{figure*}

\begin{figure}
 \includegraphics[width=\columnwidth]{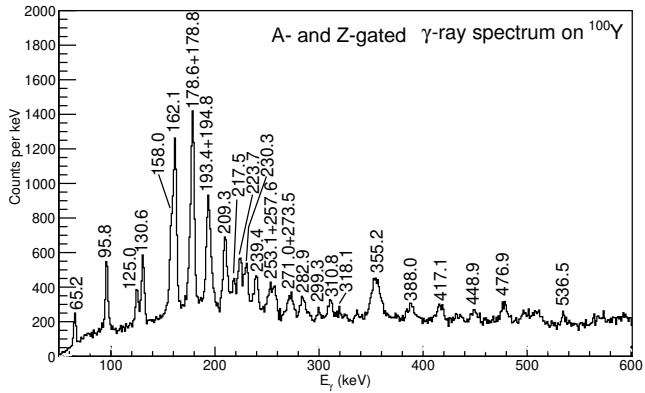}
 \caption{\label{fig2} A- and Z- gated spectrum on $^{100}$Y
 in $^{238}$U + $^{9}$Be data.}
\end{figure}

\begin{figure*}
 \includegraphics[width=0.7\textwidth]{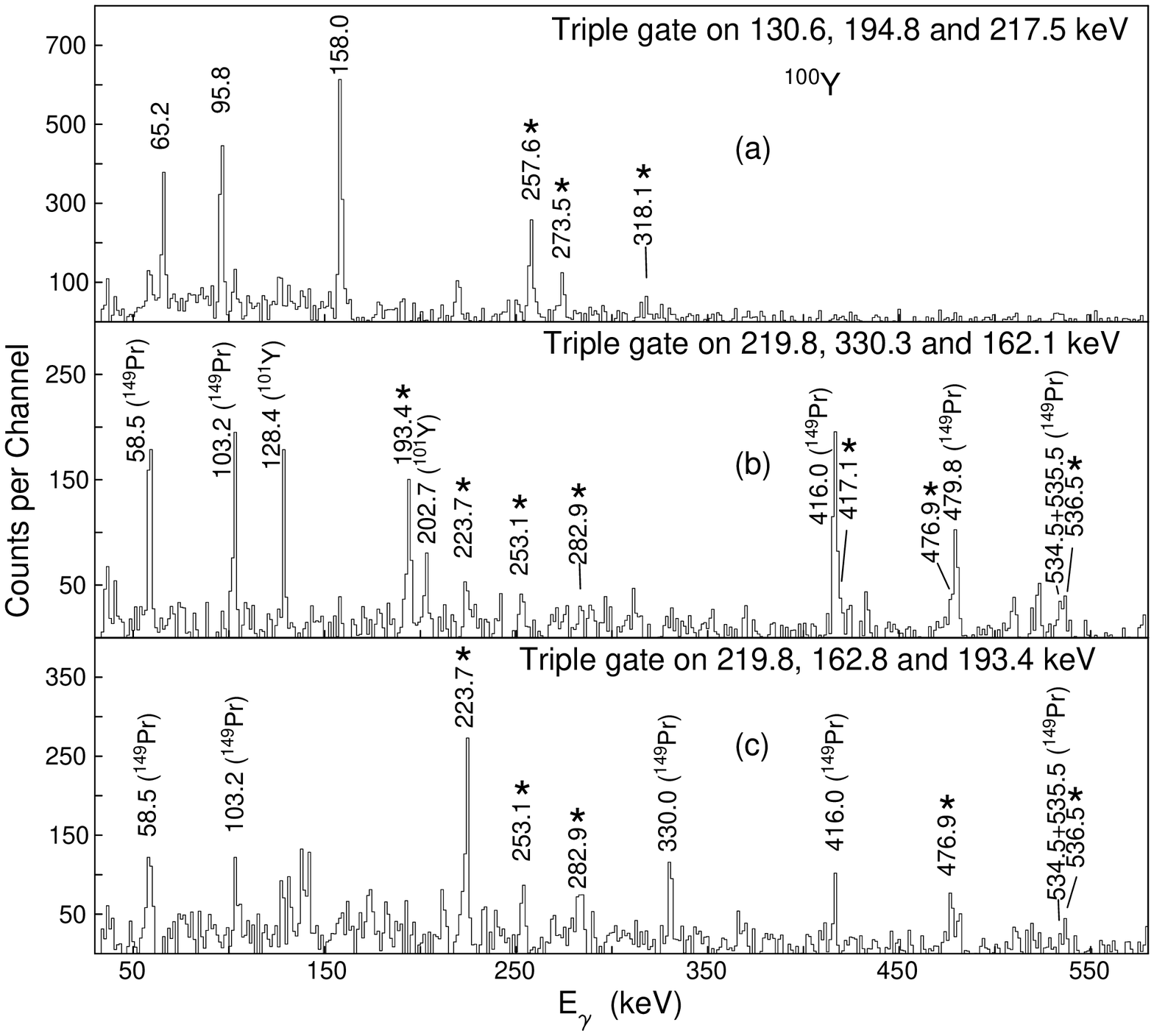}
 \caption{\label{fig3} Partial $\gamma$-ray coincidence spectrum
obtained from $^{252}$Cf spontaneous fission data. Part (a) shows a
triple gated spectrum on 130.6, 194.8 and 217.5 keV transitions in
band (1). Part (b) depicts a triple gate on 219.8 and 330.3 keV
transitions in $^{149}$Pr fission partner and 162.1 keV transitions
in $^{100}$Y. Part (c) denotes a triple gate on 219.8 keV transition
in the $^{149}$Pr fission partner and 162.1 and 193.4 keV
transitions in $^{100}$Y. Note that $^{101}$Y has a 163.3 keV
transition to bring in $^{101}$Y contamination transitions in part b).
Transitions with * are new.}
\end{figure*}

Two complementary methods have been used to investigate the level
structure of Y isotopes, which include (i) $^9$Be+$^{238}$U reaction in inverse kinematics with the unambiguous
identification of the  mass ($A$) and the proton number ($Z$) of the
emitting fission fragment, using a large acceptance spectrometer for
in-beam measurements, and (ii) the high fold $\gamma$ data from spontaneous
fission of a $^{252}$Cf source. These methods have allowed us to
identify new transitions and extend the level schemes to higher
spins of very neutron rich nuclei. In the present work the new transitions identified using
$(A,~Z)$ gated `singles' prompt $\gamma$-ray spectroscopy did not
require knowledge of the spectroscopic information of the
complementary fragment.

The first experimental measurements were
performed at GANIL using a $^{238}$U beam at 6.2 MeV/u, with an
intensity of 0.2 pnA, impinging on a 10-$\mu$m thick
$^{9}$Be target. The inverse kinematics used in
this work with forward focussed fission fragments have a
large velocity, resulting in both an efficient detection and
isotopic identification in the spectrometer. A single magnetic field
setting of the large-acceptance spectrometer VAMOS++ \cite{Rej11},
possessing a momentum acceptance of around $\pm$ 20\%, placed at
20$^\circ$ with respect to the beam axis, was used to identify
the fission fragments. The detection system (1$\times$0.15
m$^{2}$) at the focal plane of the spectrometer was composed of (i)
a Multi-Wire Parallel Plate Avalanche Counter (MWPPAC), (ii) two
Drift Chambers $(x,y)$, (iii) a Segmented Ionization Chamber
($\Delta E$), and (iv) 40 silicon detectors arranged in a wall
structure ($E_r$). The time of flight (TOF) was obtained using the
signals from the two MWPPACs, one located after the target and the
other at the focal plane (flight path $\sim$7.5 m). The parameters
measured at the focal plane [$(x,y)$, $\Delta E$, $E_r$, TOF] along
with the known magnetic field were used to determine, on an
event-by-event basis, the mass number ($A$), charge state ($q$),
atomic number ($Z$), and velocity vector after the reaction for the
detected fragment. Isotopic identifications  of elements were made
up to $Z=63$ with a mass resolution of $\Delta A/A\sim
0.4$\%~\cite{Nav13}. The prompt $\gamma$ rays were measured in
coincidence with the isotopically-identified fragments, using the
EXOGAM array~\cite{ExoGaM} consisting of 11 Compton-suppressed
segmented clover HpGe detectors placed at 15 cm from the target. The
velocity of the fragment along with the angle of the segment of the
relevant clover detector were used to obtain the $\gamma$-ray energy
in the rest frame of the emitting fragment. Errors on the
$\gamma$-ray energies of the strong transitions are 0.5~keV, while
for the weak transitions it could be as  much as 1~keV. As compared
to the results presented in Ref.~\cite{Nav13} for the Zr isotopes
the present work is the result of further improvements in the
analysis especially improving the $Z$ identification and also
involves a larger data  set.

Second, the SF of $^{252}$Cf experimental work was done by examining the
 prompt $\gamma$ rays emitted. The $\gamma$ rays were detected with the Gammasphere array at
the Lawrence Berkeley National Laboratory (LBNL). A 62 $\mu$Ci
$^{252}$Cf source was sandwiched between two iron foils of
10~mg/cm$^{2}$, which were used to stop the fission fragments and
eliminate the need for a Doppler correction. A plastic (CH) ball of
7.62 cm in diameter, surrounding the source, was used to absorb
$\beta$ rays and conversion electrons, as well as to partially
moderate and absorb fission neutrons. A total of
5.7$\times$10$^{11}$ $\gamma$-$\gamma$-$\gamma$ and higher fold
$\gamma$ events, and 1.9$\times$10$^{11}$
$\gamma$-$\gamma$-$\gamma$-$\gamma$ and higher fold $\gamma$
coincident events were recorded. These $\gamma$ coincident data were
analyzed by the RADWARE software package~\cite{Rad}. Both the triple and four fold 
coincidence data have a manual selected low energy cut from the software at $\sim$33 keV. The binning of
the triple coincidence data is about 0.67 keV/channel at the low energy end. This
value increases with the energy. The four fold coincidence data have a fixed binning
at 1.33 keV/channel. The energy calibration error for the $^{252}$Cf data is about 0.1 keV, while the fitting error for a single strong peak in the gated spectrum is usually much lower. However, in some other common cases, some weak 
or unknown contaminations in the gated spectra would enlarge the energy error which is hard to evaluate accurately. 
More experimental setup details can be found in Ref.~\cite{Ham95}.

\section{Experimental results}

\begin{figure*}
 \includegraphics[width=0.7\textwidth]{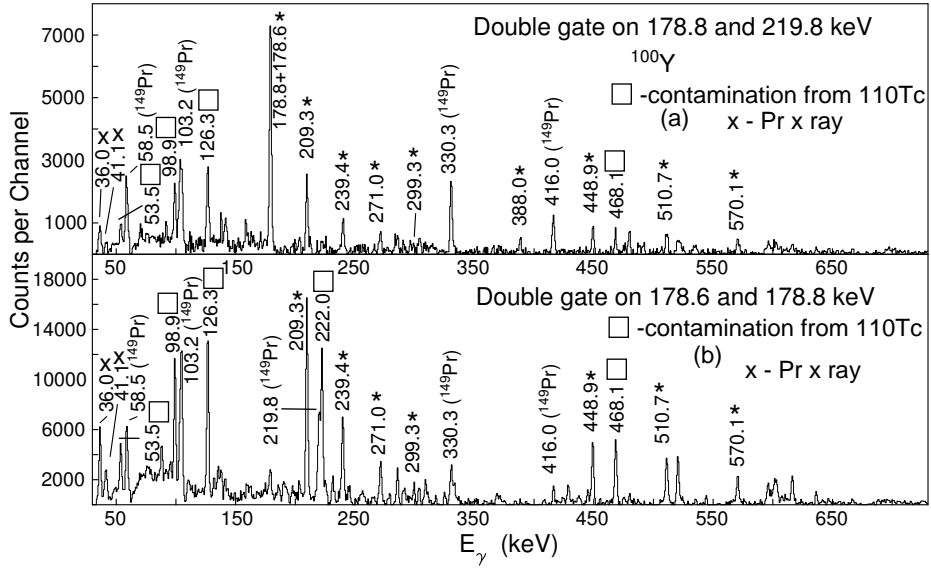}
 \caption{\label{fig5} Partial $\gamma$-ray coincidence spectrum
 obtained from $^{252}$Cf spontaneous fission data. Parts (a) and (b)
 show double gated spectra on 178.8 keV transition in $^{100}$Y and
 219.8 keV transition in $^{149}$Pr, and 178.6 and 178.8 keV
 transitions in $^{100}$Y, respectively. All the $^{100}$Y
 transitions in these two spectra are new and are labeled with an asterisk. Contamination transitions from $^{110}$Tc marked in these two spectra come from the coincidence of 222.0-178.3-178.9 keV transitions in $^{110}$Tc. Details of the level scheme of $^{110}$Tc can be found in Ref.~\cite{Luo06}.}
\end{figure*}

\begin{figure}
 \includegraphics[width=\columnwidth]{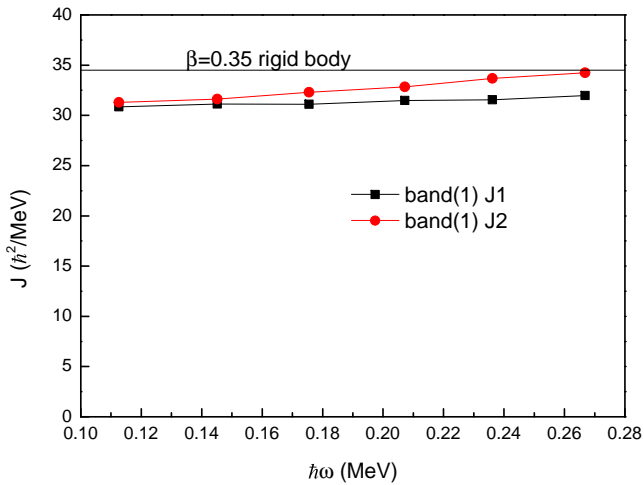}
 \caption{\label{fig6} Moments of inertia (J$^{(1)}$ and J$^{(2)}$) of band (1) in $^{100}$Y.}
\end{figure}

The level scheme of $^{100}$Y is shown in Fig.~\ref{fig1}. The
ground state of $^{100}$Y was assigned as 1$^-$ or 2$^-$ in
$\beta$-decay work~\cite{Woh}. A 1$^+$ $\pi$5/2[422] $\otimes$
$\nu$3/2[411] band was reported in Refs.~\cite{Woh,Hwa98}. In the
current work, this band (1) has been extended up to 10$^+$ with
three new transitions and levels. The new bands (2) and (3)
identified in the present work could decay to the ground state or a
145(15) keV 4$^+$ 0.94s isomeric state reported in
Refs.~\cite{Kha,Hag,Bac}. Since no transitions in bands (2) and (3) were
identified in the previous $\beta$-decay study~\cite{Woh}, these
bands are proposed to decay to the 4$^+$ isomeric state.

Fig.~\ref{fig2} shows a A- and Z- gated spectrum on $^{100}$Y
from $^{238}$U+$^{9}$Be induced fission work. In this spectrum, the
previously reported transitions in band (1) and new transitions in
bands (2) and (3) can be clearly seen. Note that the 125.0, 230.3,
310.8 keV transitions are not identified in the $^{252}$Cf SF data
so that they are not placed in the level scheme. It is worth noting
that there is a strong 125.3 keV transition in $^{99}$Y with one
neutron number less than $^{100}$Y. However, since the strong 128.4
keV transition in $^{101}$Y, the 287.2 keV transition in $^{100}$Sr,
the 212.6 keV transition in $^{100}$Zr are not clearly seen, this
spectrum is clean to $Z$ and higher mass gate. Note that the
transitions in band (2) are identical to a band reported in
$^{102}$Nb. Since the intensities of the transitions leaking through
$\Delta M$=2 gates should be very small and no such leakage is seen
in any other $A/Z$ gated spectra in the previous work, the
transitions in band (2) in the $^{100}$Y A- and Z- gated spectrum
belong to $^{100}$Y rather than $^{102}$Nb.

Fig.~\ref{fig3} (a) shows a $\gamma$ ray coincidence spectrum by
triple gating on three known transitions in band (1) from $^{252}$Cf
spontaneous fission data. The 257.6, 273.5 and 318.1 keV new
transitions in band (1) can be seen. Fig.~\ref{fig3} (b) shows a
$\gamma$ ray coincidence spectrum by gating on two transitions in
the strongly populated 3n fission partner $^{149}$Pr and the strong
162.1 keV transition identified in the A/Z gated spectrum in
Fig.~\ref{fig2}. Despite the contamination transitions in $^{101}$Y
(the 2n partner of $^{149}$Pr), which originate from the 163.3 keV
transition in $^{101}$Y, the 193.4, 223.7, 253.1, 282.9 keV new
transitions identified in the A/Z gated $^{100}$Y spectrum are
clearly seen. The 223.7, 253.1, 282.9, 476.9 and 536.5 keV transitions in
band (2) are also seen in the coincidence spectrum gating on a
strong transition in $^{149}$Pr and the 162.1 and 193.4 keV
transitions in Fig.~\ref{fig3} (c). These coincidence spectra from
$^{252}$Cf SF further confirm that the somewhat broadened peak
around 160 keV in Fig.~\ref{fig2} contains two components of 158 and
162 keV. Fig.~\ref{fig5} (a) shows a $\gamma$ ray coincidence spectrum
by double gating on a strong transition in band (3) of $^{100}$Y and
a $^{149}$Pr fission partner transition. The correlated 179, 209,
239, 271, 299, 388, 449, 511 and 570 keV transitions
in band (3) can be seen. These transitions are also
confirmed in the 178.6 and 178.8 keV double coincidence spectrum in
Fig.~\ref{fig5} (b).

\section{Discussion}

Band (1) was assigned to have a \linebreak 1$^+$ $\pi 5/2[422] \otimes \nu
3/2[411]$ configuration~\cite{Woh,Hwa98}. Such configuration is
adopted in the current work. The neutron pairing gap and proton
pairing gap of this state are proposed to drop down~\cite{Woh},
which result in the moments of inertia very close to a rigid body
one (Fig.~\ref{fig6}). Note that the M1 transition energies of the
same spins in bands (1), (2) and (3) are very close, e.g. 158 and
162 keV from 5$^+$ to 4$^+$, 195, 193 and one of the 179 keV from 6$^+$ to
5$^+$. Therefore, the moments of inertias in these two bands are
also very close to rigid body. However, only band (1) is shown in
Fig.~\ref{fig6} because the values in band (2) and (3) would be too
close to be seen in a figure.

Experimental measurements on $Z=36$-$40$ isotopes revealed a sudden
onset of deformation at $N \approx 60$. For yttrium isotopes,
deformation was small for \linebreak $N=48$-$58$ and shape coexistence was
reported for $N=59$~\cite{Bac}. Then, the deformation suddenly
changes to a higher value ($\beta_2$ $\sim$ 0.4) for $N=60$, $62$ with
almost no triaxiality. Cheal~\emph{et al.} reported a tentative spin
3 for the ground state of $^{100}$Y with 0.39(4) $\beta_2$
deformation from laser spectroscopy~\cite{Che}.
Baczynska~\emph{et al.} reassigned this state as a 0.35(4) $\beta_2$
deformation, spin 4 isomeric state by using a similar experimental
approach~\cite{Bac}. A probable 4$^{+}$ $\pi 5/2[422] \otimes \nu
3/2[411]$ configuration was suggested from comparison of $g$ factors
between experiment and theory~\cite{Bac}. Similar and regular energy
spacing with no signature splitting in bands (2) and (3) indicate a
large rigid deformation, which is consistent with systematics. Note
that $^{99,101}$Y are both proposed to have large quadrupole
deformation~\cite{Bac,Che,Luo05}. As discussed in the previous
section, the band-head of band (2) is proposed to be the 145(15) keV
4$^{+}$ isomer. The configurations of these bands would be 4$^{+}$
$\pi 5/2[422] \otimes \nu 3/2[411]$ and 5$^{+}$ $\pi 5/2[422]
\otimes \nu 5/2[413]$, respectively, according to the Nilsson
orbitals in this region. Bandheads of (2) and (3) would form a $\pi
5/2[422] \otimes \nu [312$ $5/2,3/2]$ neutron pseudo spin doublet.
Bands (1) and (2) would form a GM doublet and originate from the
Gallagher Moskowski interaction, in which the intrinsic spins of
proton and neutron are coupled in parallel and antiparallel to get
low $K=|\Omega_\pi-\Omega_\nu|$ and high $K=\Omega_\pi+\Omega_\nu$.
In the neighboring $^{98}$Sr and $^{99}$Y isotopes, this GM
structure has also been observed~\cite{Li,Mey,Luo05}.

\subsection{Constrained CDFT calculations}

In order to understand the nature of the observed band structure in
$^{100}$Y, constrained triaxial covariant density functional theory
(CDFT) calculations~\cite{J.Meng2006PRC, Ring1996PPNP,
Vretenar2005PR, J.Meng2006PPNP, J.Meng2011PIP, J.Meng2016book} were
first performed to obtain the potential energy surface (PES) to
search for the possible configurations and deformation parameters.
Subsequently, with the obtained configurations and deformations,
quantal particle rotor model (PRM)~\cite{Frauendorf1997NPA,
J.Peng2003PRC, S.Q.Zhang2007PRC, B.Qi2009PLB, Q.B.Chen2010PRC,
B.Qi2011PRC, Ayangeakaa2013PRL, Lieder2014PRL, Kuti2014PRL}
calculations were carried out to reproduce the energy spectra and
electromagnetic transition probabilities and investigate the angular
momentum geometries for bands (1), (2), and (3).

In the constrained triaxial CDFT calculations, the point-coupling
energy density functional PC-PK1~\cite{P.W.Zhao2010PRC} in the
particle-hole channel is adopted, while the pairing correlations in
the particle-particle channel are neglected. The neglecting of the
pairing correlations in the CDFT calculations is consistent with the
fact that the extracted experimental moments of inertia of the three
bands are very close to a rigid body one as shown in
Fig.~\ref{fig6} and the discussion above. The solutions of the equation of motion for the
nucleons are accomplished by an expansion of the Dirac spinors in a
set of three-dimensional harmonic oscillator basis functions in
Cartesian coordinates with 12 major shells. The PES is obtained by
constraining the $(\beta_2,\gamma)$ deformation parameters in the
intervals $\beta_2 \in[0.0,0.8]$ and $\gamma\in [0^\circ,60^\circ]$,
with step sizes $\Delta \beta_2=0.05$ and $\Delta\gamma=6^\circ$,
respectively.

\begin{figure}[!ht]
  \begin{center}
  \hspace{-2.5 cm}
     \includegraphics[width=8.5 cm]{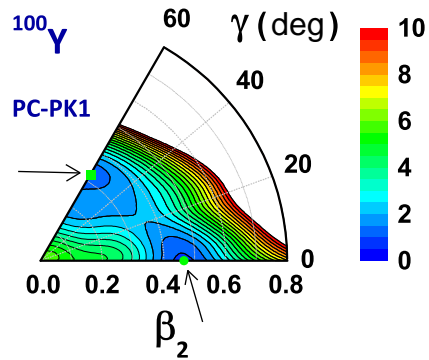}
  \vspace{-2.2 cm}
        \caption{(Color online) Potential energy surface in the
        $\beta_2$-$\gamma$ plane $(0\leq \beta_2 \leq 0.8, 0\leq \gamma\leq 60^\circ)$
        for the ground-state configuration of $^{100}$Y in constrained triaxial CDFT
        calculations with the PC-PK1 effective interaction. All energies are normalized
        with respect to the energy of the absolute minimum (in MeV) indicated by the dot and arrow.
        The energy separation between each contour line is 0.5 MeV.}\label{fig7}
  \end{center}
\end{figure}

In Fig.~\ref{fig7}, the PES in the $\beta_2$-$\gamma$ plane for
$^{100}$Y calculated by the constrained triaxial CDFT is shown. It
shows that the minimum of the PES (ground state, labeled by the dot at the bottom)
locates at ($\beta_2= 0.45$, $\gamma=0^\circ$), which corresponds to
a large prolate shape. Meanwhile, it is interesting to note that
there is a local minimum (labeled by the square on the left) around at
($\beta_2= 0.32$, $\gamma=60^\circ$). However, its energy is higher
than that of the ground state by about 1.0 MeV; even this value is much
larger than that of the bandhead energy of band (3), \linebreak 334 keV. Thus it
can be understood that the band built on this oblate state has not
been observed in the present experiment.

It is found that the valence nucleon configuration of the ground
state of $^{100}$Y is expressed as spherical orbits,
$\pi(1g_{9/2})^5 \otimes\nu (2d_{5/2})^3(1g_{7/2})^4(1h_{11/2})^4 $,
which has a positive parity. Considering that the $2d_{5/2}$ and
$1g_{7/2}$ orbitals are a pair of pseudo partners with too
large admixtures to be distinguished, and the paired orbits have no
contributions to the total angular momentum, the ground state
configuration can be rewritten as $\pi(1g_{9/2})^1 \otimes\nu
(2d_{5/2})^1$ or $\pi(1g_{9/2})^1 \otimes\nu (1g_{7/2})^{-1}$. The
former corresponds to the configuration $\pi 5/2[422]\otimes \nu
3/2[411]$ and will be assigned to bands (1) and (2), and the latter $\pi 5/2[422]\otimes \nu 5/2[413]$ to band (3). Such
configuration assignments are consistent with the previous
investigations in Refs.~\cite{Woh,Hwa98,Bac} and the systematic
analyses discussed above, and will be examined and confirmed by
the following comparisons between the PRM calculated results and the
experimental observations.

\subsection{PRM calculations}

With the configurations and deformation parameters obtained from the
constrained triaxial CDFT calculations, it is straightforward to
perform the PRM calculations to study the energy spectra, the
electromagnetic transition probabilities, and the angular momentum
geometries for the observed bands (1), (2), and (3) in $^{100}$Y.

\begin{figure}[!ht]
  \begin{center}
     \includegraphics[width=8 cm]{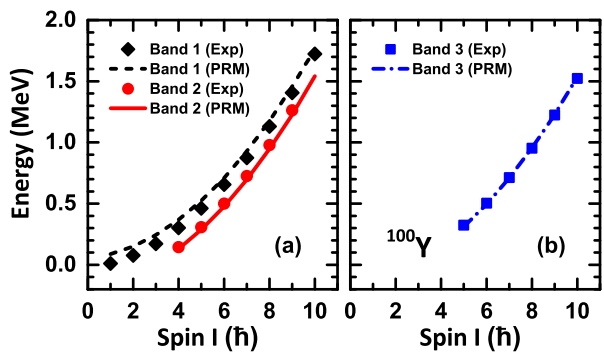}
        \caption{(Color online) The energy spectra for bands (1), (2), and (3) calculated
        by PRM in comparison with the data.}\label{fig8}
  \end{center}
\end{figure}

The many-particle-many-hole PRM calculations have been carried out
for bands (1) and (2) with the configurations $\pi(1g_{9/2})^5
\otimes\nu (2d_{5/2})^3$, and for band (3) with $\pi(1g_{9/2})^5
\otimes\nu (1g_{7/2})^5$. In both calculations, the deformation
parameters are used as ($\beta_2=0.45$, $\gamma=0^\circ$), obtained
from the constrained triaxial CDFT calculations. With the
deformation parameters, the single-$j$ shell Hamiltonian coupling
parameter can be calculated by~\cite{S.Y.Wang2009CPL}
\begin{align} C=\Big(\frac{123}{8}\sqrt{\frac{5}{\pi}}\Big)
\frac{2N+3}{j(j+1)}A^{-1/3}\beta_2,
\end{align}
where $N$ denotes the number of oscillator quanta. The moments of
inertia are chosen to be $\mathcal{J}=20.0~\hbar^2/\rm {MeV}$. In
the electromagnetic transition probabilities calculation, the
empirical intrinsic quadrupole moment \linebreak
$Q_0=(3/\sqrt{5\pi})R_0^2Z\beta_2$ with $R_0=1.2A^{1/3}~\rm fm$, the
gyromagnetic ratio $g_R=Z/A=0.39$, $g_{\pi}(g_{9/2})=1.26$,
$g_{\nu}(d_{5/2})=-0.46$, and $g_{\nu}(g_{7/2})=0.70$ are
adopted~\cite{Ring1980book}.

\begin{figure}[!ht]
  \begin{center}
     \includegraphics[width=8 cm]{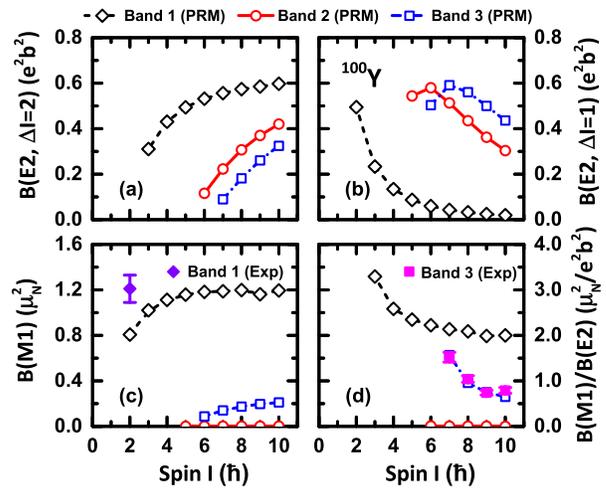}
        \caption{(Color online) The in-band reduced electromagnetic transition
        probabilities $B(E2, \Delta I=2)$, $B(E2, \Delta I=1)$, and $B(M1)$, and
        the $B(M1)/B(E2)$ ratios of bands (1), (2), and (3) calculated by the PRM
        in comparison with the available data. The experimental $B(M1)$ value for the 2$^+$ state in band (1) is extracted from the $\beta$ decay data \cite{Mach1990PRC}, as shown in the left bottom part (c) for comparison. The experimental value for $B(M1)/B(E2)$ value obtained in the present work for band (3) is also included in part (d) for comparison. }\label{fig9}
  \end{center}
\end{figure}

\begin{figure*}[!ht]
  \begin{center}
     \includegraphics[width=12 cm]{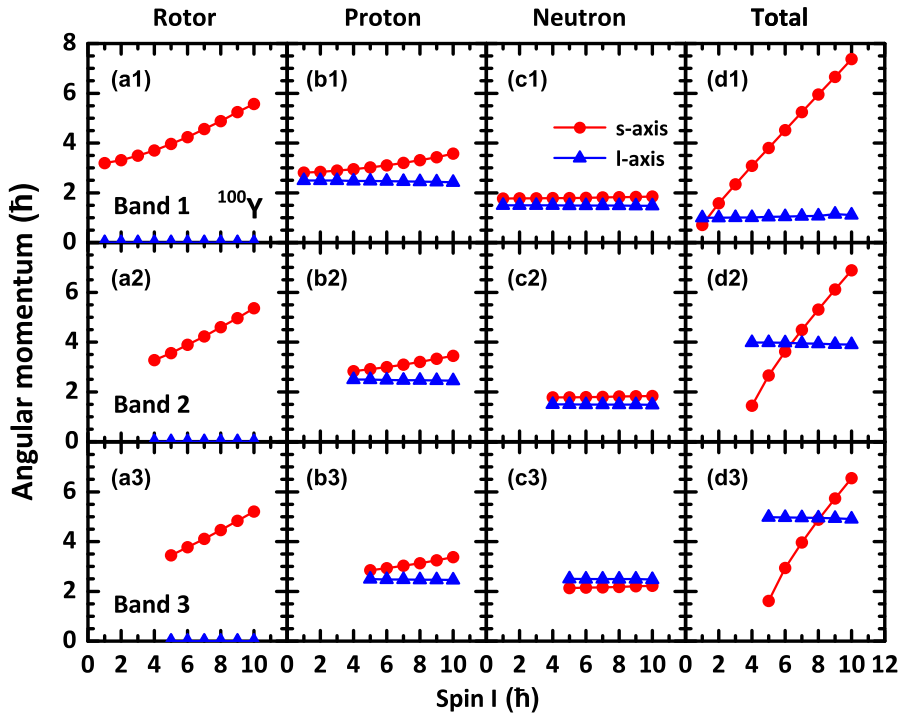}
        \caption{(Color online) The root mean square components of the angular momentum along the
        short ($s$-, circles) and long ($l$-, triangles) axes of the
        core ($\bm{R}$), the proton ($\bm{j_\pi}$), the neutron ($\bm{j_\nu}$),
        and the total spin ($\bm{I}$) calculated by PRM for bands (1),
        (2), and (3).}\label{fig10}
  \end{center}
\end{figure*}

In Fig.~\ref{fig8}, the energy spectra for bands (1), (2), and (3)
calculated by PRM are shown in comparison with the data. It is seen
that the PRM results excellently agree with the data, which confirms
the configuration assignments for these three bands. For bands (1)
and (2), they are separated by a nearly constant value ($\sim
150~\rm{keV}$) over the whole spin region. This can be understood, in
the following, that they are built on the different bandheads, that
the intrinsic spins of proton and neutron are coupled in parallel
and antiparallel to low $K=|\Omega_\pi-\Omega_\nu|=1\hbar$ and high
$K=\Omega_\pi+\Omega_\nu=4\hbar$. For bands (2) and (3), their
energy differences are rather small (within $20~\rm{keV}$). This is
due that they are built on a pair of nearly degenerated pseudo spin
partner configurations $\nu(2d_{5/2})$ and $\nu (1g_{7/2})$. Here,
we should mention that the present PRM cannot give this energy
difference since the two calculations are carried out separately
with different model spaces.

The in-band reduced electromagnetic transition probabilities $B(E2,
\Delta I=2)$, $B(E2, \Delta I=1)$, and $B(M1)$, and the
$B(M1)/B(E2)$ ratios of bands (1), (2), and (3) calculated by the
PRM are presented in Fig.~\ref{fig9}, together with the available
data. The $B(M1)$ value in band (1) extracted from beta decay \cite{Mach1990PRC} and the $B(M1)/B(E2)$ ratios extracted from the current work assume pure M1 for the $\Delta I=1$ transitions. Because of the identical transitions in $^{102}$Nb \cite{Hwa98} and the 416, 479, 535 keV transitions in the $^{149}$Pr fission partner, it is difficult to obtain accurate transition intensities and $B(M1)/B(E2)$ ratios in band (2). Thus, such ratios in band (2) from experiment data are not provided in the present work.
For the $B(E2, \Delta I=2)$, it increases smoothly with spin.
It is seen that it becomes smaller with the larger $K$, which is due
to the Clebsch-Gordon coefficients \linebreak $\langle I_iK20|I_fK\rangle$
($I_i=I_f+2$)~\cite{Bohr1975}. For the $B(E2, \Delta I=1)$, contrary
to $B(E2, \Delta I=2)$, it decreases with spin. Moreover, it shows
strong competitions with $B(E2, \Delta I=2)$. Large $B(E2, \Delta
I=1)$ is accompanied by small $B(E2, \Delta I=2)$, vice versa.

For the $B(M1)$, it is seen that the band (1) shows a large value
($\sim 1.2~\mu_N^2$). This result is in good agreement with the
experiment value of $2^+$ state, as seen in Fig.~\ref{fig9}(c), from
$\beta$ decay measurement~\cite{Mach1990PRC}. This further supports
the configuration assignment for band (1). However, one notes that
its partner band (2) presents an almost vanished value. The reason
can be understood from the value of the $g$-factor and the coupling
modes in these two bands. For the $g$-factor, we have
$g_{\pi}(g_{9/2})-g_R=0.87$ and $g_{\nu}(d_{5/2})-g_R=-0.85$. As a
consequence, for the band (1), where proton and neutron are coupled
in antiparallel, the $B(M1)$ values are enhanced. In contrast, for
band (2) with the proton and neutron being parallel coupled, the
$B(M1)$ values are reduced significantly. Similarly, as
$g_{\nu}(g_{7/2})-g_R=0.31$ is small. The $B(M1)$ values of band (3)
are small as well. Here, it is worth mentioning that, according to
the PRM results, the in-band $\Delta I=1$ transitions in bands (2)
and (3) are dominated primarily by $E2$ transitions rather than $M1$
transitions. Therefore, it is highly expected that the angular
distribution and the polarization asymmetry would be measured for
these transitions in the future experiments to identify their
multipolarities.

For the $B(M1)/B(E2)$ ratio, it decreases with spin for bands (1)
and (3), and almost vanishes for band (2). The former one is caused
by the increase $B(E2)$, and the latter one by the vanished $B(M1)$,
as seen in Figs.~\ref{fig9}(a) and (c). The available experimental
values of band (3) are reproduced by the PRM very well. This is
further in favor of the configuration assignment for band (3).

The successes in reproducing the energy spectra and electromagnetic
transition probabilities of the bands (1), (2), and (3) in $^{100}$Y
by PRM encourage us to investigate the corresponding angular
momentum geometries to learn the coupling modes in these three
bands. In Fig.~\ref{fig10}, the root mean square components along
the short \linebreak ($s$-, circles) and long ($l$-, triangles) axes of the
core ($\bm{R}$), the proton ($\bm{j_\pi}$), the neutron
($\bm{j_\nu}$), and the total spin ($\bm{I}$) calculated by PRM for
bands (1), (2), and (3) are thereby illustrated. Note that since the
three bands are all prolate deformed, here the $l$-axis is the
symmetry axis, and the $s$-axis is the axis that is perpendicular to
the symmetry axis.

For the rotor, its rotational angular momentum aligns along the $s$-axis for all three bands, and
its $l$-component vanishes as expected in axially deformed bands. As
the total spin increases, $R_s$ also increases gradually. For the
$1g_{9/2}$ valence protons, they mainly align along the $s$-axis.
Its $l$-component keeps almost a constant value with \linebreak
$\Omega_\pi=5/2\hbar$. For the $2d_{5/2}$ valence neutrons in bands
(1) and (2) and the $1g_{7/2}$ valence neutrons in band (3), the
three components are very similar, and are almost constant. It is
noted that the $l$-component is $\Omega_\nu=3/2\hbar$ in the former
ones, and $\Omega_\nu=5/2\hbar$ in the latter one. For the total
spin, its $s$-components linearly increase with spin, which is
induced by the increases of the rotor and the valence protons. It is
interesting to observe that its $l$-component is quite different in
the three bands. Namely, $K=1$, $4$, and $5\hbar$ in the bands 1, 2,
and 3, respectively.

Therefore, according to the above angular momentum geometry
analyses, it is clearly seen that the protons and neutrons in the
bands (2) and (3) are coupled in parallel to respectively get
$K=\Omega_\pi+\Omega_\nu=4$ and $5\hbar$, while they are
antiparallel to get $K=|\Omega_\pi-\Omega_\nu|=1\hbar$ in the band
(1). This strongly supports the GM doublet interpretation for bands
(1) and (2), and the pseudo spin doublet interpretation for bands
(2) and (3). This observation provides solid experimental
evidence for the coexistence of the GM doublet and pseudo spin
doublet bands in the same nucleus.

\section{Summary}

In summary, the new transitions and levels in neutron rich $^{100}$Y have been
identified by $^9$Be+$^{238}$U experiment with A- and Z- gate with fragment identification. The coincidence of these transitions have been investigated by analyzing the
high statistics $\gamma$-$\gamma$-$\gamma$ and
$\gamma$-$\gamma$-$\gamma$-$\gamma$ coincidence data from the
spontaneous fission of $^{252}$Cf.
Two new bands have been identified for the first time built on the isomer. The configurations of the band structures in $^{100}$Y have been assigned from the theoretical calculations.

Constrained triaxial CDFT calculations and quantal PRM calculations
have been performed to explain the observed band structures. The
obtained PES from the CDFT calculations suggests that the $^{100}$Y
has a large prolate deformed ground state with $\beta_2=0.45$. With
the obtained deformation parameter and the configuration, the PRM
calculations excellently reproduce the experimental energy spectra
and the available electromagnetic transition probabilities for bands
(1), (2), and (3). By analyzing the angular momentum geometries, it
is confirmed that the $K=4^+$ new band is a high-$K$ GM partner of
the known $K=1^+$ band, with the $\pi 5/2[522] \otimes \nu 3/2[411]$
configuration, and is also proposed to be the pseudo spin partner of
the new $K=5^+$ band with a $\pi 5/2[422] \otimes \nu 5/2[413]$
configuration, to form a $\pi 5/2[422] \otimes \nu [312$ $5/2,3/2]$
neutron pseudospin doublet. This observation provides solid
experimental evidence for the coexistence of the GM doublet and
pseudo spin doublet bands in a same nucleus.

\begin{acknowledgments}

The work at Vanderbilt University and Lawrence Berkeley National
Laboratory are supported by the US Department of Energy under Grant
No. DE-FG05-88ER40407 and Contract No. DE-AC03-76SF00098. The work
at Tsinghua University was supported by the National Natural Science
Foundation of China under Grant No. 11175095. The work at  JINR was
supported  by   the  Russian  Foundation  for   Basic  Research
Grant No.  08-02-00089 and  by the  INTAS Grant  No. 03-51-4496. The
work at Peking University are supported by the Chinese Major State
973 Program No. 2013CB834400, the National Natural Science
Foundation of China (Grants No. 11335002, No. 11375015, and No.
11461141002), and the China Postdoctoral Science Foundation under
Grants No. 2015M580007 and No. 2016T90007. One of us (S.B.)
acknowledges partial financial support through the  LIA France-India
agreement. We would like to thank J.~Goupil, G.~Fremont,
L.~M\'{e}nager, J.~Ropert, C.~Spitaels, and the  GANIL accelerator
staff for their technical contributions.

\end{acknowledgments}

\end{document}